# Exchange Bias and Bistable Magneto-Resistance States in Amorphous TbFeCo thin Films

*Authors:*

Xiaopu Li,[1,a] Chung T. Ma,[1] Jiwei Lu,[2] Arun Devaraj,[3] Steven R. Spurgeon,[4] Ryan B. Comes,[4] and S. Joseph Poon[1,b]

[1]Department of Physics and [2]Department of Materials Science and Engineering, University of Virginia, Charlottesville, Virginia 22904, USA

[3]Environmental Molecular Sciences Laboratory and [4]Physical and Computational Sciences Directorate, Pacific Northwest National Laboratory, Richland, WA 99352, USA

*Abstract:*

Amorphous TbFeCo thin films sputter deposited at room temperature on thermally oxidized Si substrate are found to exhibit strong perpendicular magnetic anisotropy (PMA). Atom probe tomography (APT), scanning transmission electron microscopy (STEM), and energy dispersive spectroscopy (EDS) mapping have revealed two nanoscale amorphous phases with different Tb atomic percentages distributed within the amorphous film. Exchange bias accompanied by bistable magneto-resistance states has been uncovered near room temperature by magnetization and magneto-transport measurements. The exchange anisotropy originates from the exchange interaction between the ferrimagnetic and ferromagnetic components corresponding to the two amorphous phases. This study provides a platform for exchange bias and magneto-resistance switching using single-layer amorphous ferrimagnetic thin films that require no epitaxial growth.

[a] Email: xl6ba@virginia.edu

[b] Email: sjp9x@virginia.edu

*Main Content:*

The exchange bias (EB) effect describes a unidirectional shift of a magnetic hysteresis loop along the magnetic field axis.[1, 2] Recently the EB effect has received intensive study because of its importance in a variety of technological applications, especially in spin-valve devices and magnetic tunnel junctions.[3-9] The exchange anisotropy has been interpreted in terms of the exchange interaction across the ferromagnetic (FM)-antiferromagnetic (AFM) interface, e.g. in Co/IrMn.[10-12] EB also exists in soft/hard FM/FM systems, where two types of EB have been reported, i.e. minor loop effect and standard EB effect.[13-14] Moreover, enhanced EB has been observed using compensated ferrimagnetic (FiM) materials, e.g. $GdCo_2$-Co and TbFe-[Co/Pt].[15-16] Recently, polycrystalline Heusler compounds Ni-Mn-$X$ ($X$ = Sn, In, Sb) and Mn-Pt-Ga have been reported to show an intrinsic EB at low temperature due to the coexistence of FM and AFM regions.[6, 7, 17, 18] However, ongoing efforts seek to engineer more desirable properties, such as high Néel temperature, large magnetic anisotropy and good chemical and structural tunability. In this work, we present the amorphous rare-earth-transition-metal (RE-TM) thin film as one promising material-base that provides a wide compositional tunability and requires no epitaxial growth.

Amorphous RE-TM thin films with perpendicular magnetic anisotropy (PMA) are currently under investigation for their applications in high-density low-current spintronic devices and ultrafast magnetic switching.[19-22] Amorphous TbFeCo possesses strong PMA with $K_u \sim 5 \times 10^6 \text{erg/cm}^3$, and is FiM containing AFM coupled Tb and FeCo sublattices.[5, 23] These compounds display a compensation temperature ($T_{comp}$) for a range of Tb concentrations.[22, 24] This compensation phenomenon can be understood in terms of the conventional FiM model, where antiparallel sublattice moments compensate for each other to produce zero net magnetic moment.[25] Additionally, anisotropic microstructures have been reported in room-temperature sputter-deposited thin films.[26-28] Of special note are compositional inhomogeneities that are formed by a shadowing effect if separate sources are used for multi-target cosputter deposition. These growth-induced inhomogeneities provide an opportunity to manipulate the magnetic properties of the amorphous RE-TM thin film. In this letter, EB and bistable magneto-resistance (MR) states have been uncovered at room temperature in the amorphous TbFeCo thin films. Two growth-induced nanoscale phases have been observed by atom probe tomography (APT) and aberration-corrected scanning transmission electron microscopy (STEM). The EB originates from the exchange interaction between the FiM and FM components corresponding to the two nanoscale phases respectively. The bistable MR states can also be understood in light of the same exchange interaction.

Amorphous $Tb_{26}Fe_{64}Co_{10}$ (a-TbFeCo) thin films were prepared on thermally oxidized Si substrates by RF magnetron sputtering at room temperature. A series of films were made with thickness from 50 to 200 nm. Since the films with different thicknesses exhibit similar magnetic features, only the results of the 100-nm thick films are shown herein. The samples were capped by a 5-nm-thick Ta layer to prevent oxidation. A typical pattern of amorphous structure was provided by cross-sectional high resolution transmission electron microscopy (HRTEM), indicating no apparent crystallinity, similar to that reported previously.[23, 29] The experimental details and HRTEM results are provided in the supplementary material.[30]

To characterize the compositional uniformity, we have conducted high-angle annular dark field imaging (STEM-HAADF). Figure 1(a) shows a representative STEM-HAADF micrograph. The non-uniform contrast of the image indicates local compositional fluctuations. To further validate the STEM-HAADF observations we utilized energy-dispersive X-ray spectroscopy (STEM-EDS). Figure 1(b-e) shows the HAADF signal and maps of the Co *K*, Tb *L*, and Fe *K* edges taken around one such cluster. These maps have had their background removed and overlapping edges deconvoluted. Qualitatively it is clear that the distribution of all three signals is non-uniform. Focusing specifically on the Tb *L* and Fe *K* edges, we see that in the regions marked with arrows there is a local depletion in Tb that directly coincides with an enrichment in Fe. This suggests that the distribution of these two elements is inversely related. A composite map of the Tb *L* and Fe *K* edges (Figure 1(f)) shows this even more clearly and supports the conclusion that local Fe enrichment is associated with local Tb depletion on the length scale of 2-5 nm.

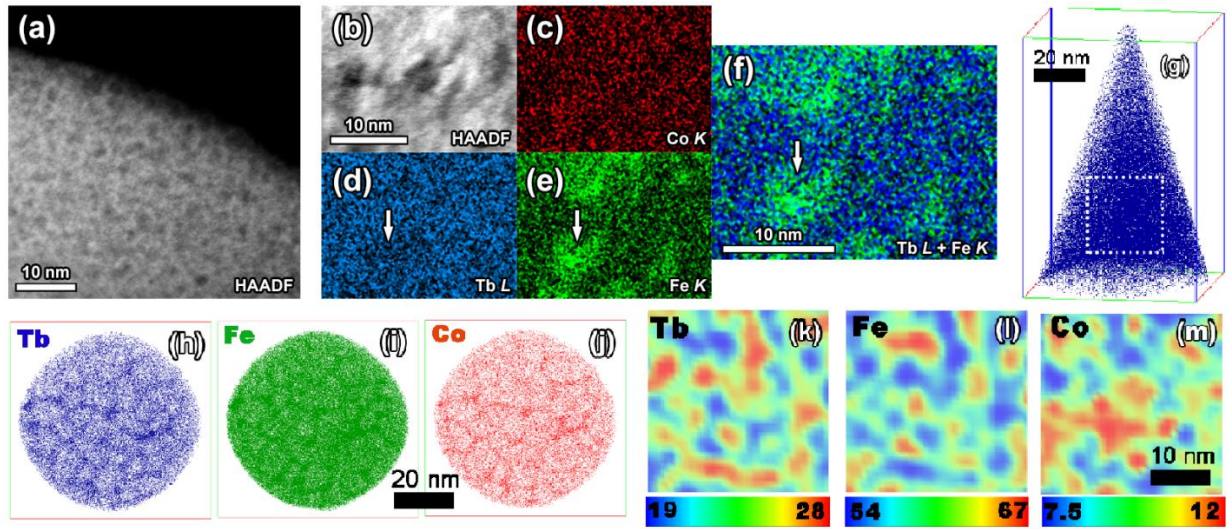

**Figure 1. Correlative STEM and APT analysis.** a) Representative STEM-HAADF micrograph exhibiting non-uniform contrast due to clustering. b-e) STEM-EDS maps of the HAADF, Co $K$, Tb $L$, and Fe $K$ signals, respectively, around one such cluster. f) Composite of the Tb $L$ and Fe $K$ edges. g) Tb (blue) distribution in the $67.66 \times 66.13 \times 99.89$ nm volume analyzed by APT. h-j) 5-nm slice of APT data perpendicular to z axis showing Tb (h), Fe (i) and Co (j) distribution parallel to the film plane. k-m) 2D concentration maps of Tb (k), Fe (l) and Co (m) plotted on a $1 \times 30 \times 30$ nm volume shown by the dashed rectangle in (g). The dark red and dark blue show the highest and lowest concentration regions respectively. The scale bar indicates the corresponding high and low concentrations for each map.

APT provides information on 3D nanoscale distribution of elements in amorphous thin films, permitting quantitative measurements of uniformity of elemental distribution, which can complement the qualitative observations by 2D STEM-EDS mapping.[31-36] APT analyzed a 3D volume of $67.66 \times 66.13 \times 99.89$ nm and Figure 1(g) shows its Tb distribution. Tb (blue), Fe (green) and Co (red) distribution along a 5-nm slice parallel to the film plane is shown in Figure 1(h-j). These clearly show a continuous network-like segregation of all three elements. A $1 \times 30 \times 30$ nm volume region highlighted by the dotted rectangle was selected to plot 2D concentration elemental maps to obtain a quantitative distribution. In all three 2D maps provided in Figure 1(k-m) red color indicates the highest concentration regions and blue indicates the lowest. It is clearly observable that Tb segregates to distinct regions which are depleted in Fe and Co. This observation directly correlates with the STEM-EDS measurements, supporting the existence of a two-phase compositional partitioning in the a-TbFeCo thin film.

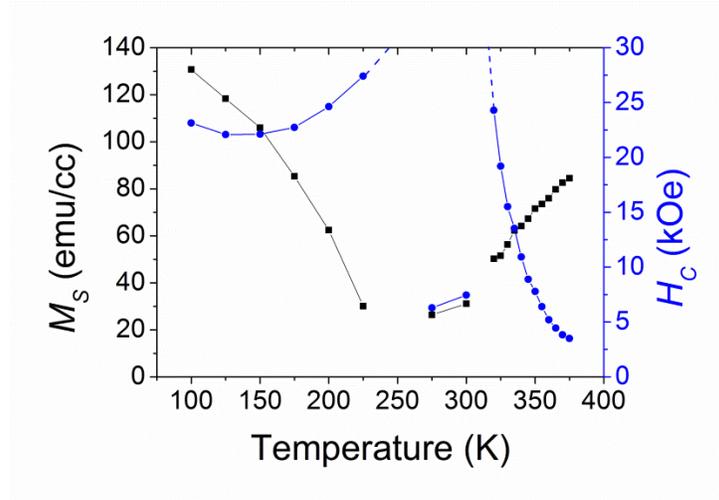

**Figure 2. Temperature dependence of $M_S$ (black) and $H_C$ (blue) of the amorphous $Tb_{26}Fe_{64}Co_{10}$ thin film.** The reduced $H_C$ at 275 and 300 K is related to the exchange bias. Hysteresis loops at 300 K are provided in Figure 3(a).

The a-TbFeCo films exhibit strong PMA for a wide range of temperatures. The magnetic hysteresis loops were characterized by vibrating sample magnetometry (VSM) as a function of temperature from 100 to 375 K, from which the temperature dependence of saturation magnetization ($M_S$) and coercivity ($H_C$) were extracted. As shown in Figure 2, the Curie temperature ($T_C$) of the system is greater than 375 K. $M_S$ is expected to reach a minimum at $T_{comp}$, which is near 250 K.

Figure 3(a) shows hysteresis loops in the out-of-plane direction for three distinct temperature regions: well-below, near, and well-above $T_{comp}$. Two exchange biased hysteresis loops have been observed at 300 K in the region near $T_{comp}$. Both of the biased loops have the absolute EB field ($|H_E|$) of 1.9 kOe. The observed EB can be either positive or negative depending on the sample initialization condition. The biased loop with negative $H_E$ is initialized by heating the sample to 355 K in zero field, followed by magnetizing it in +30 kOe and finally cooling down to the original temperature of 300 K in zero field. On the other hand, the biased loop with positive $H_E$ is initialized by first cooling the sample down to 175 K in zero field, followed by magnetizing it in +30 kOe and finally heating up to 300 K in zero field. The EB vanishes at temperature well-below and well-above $T_{comp}$, namely 175 and 355 K.

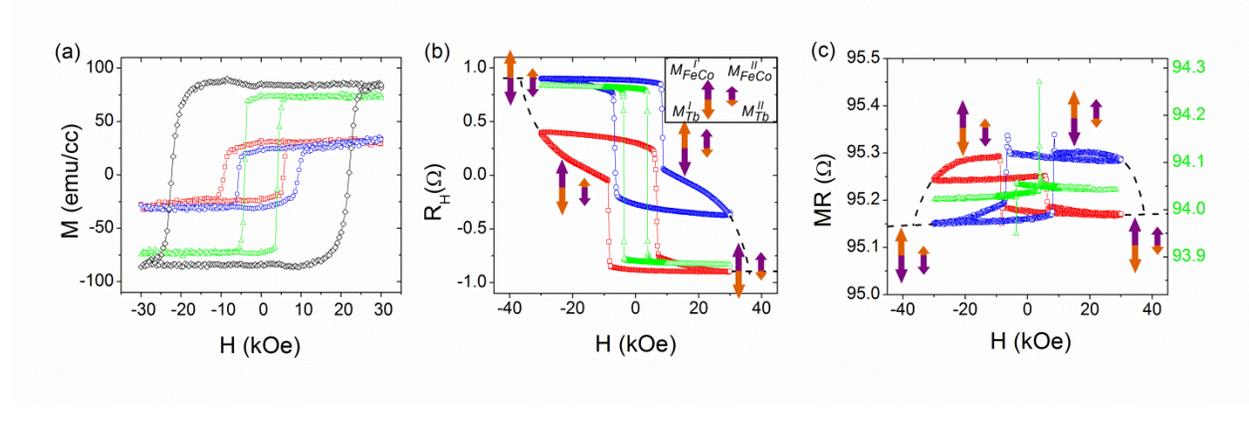

**Figure 3. Magnetic and magneto-transport measurements of the amorphous $Tb_{26}Fe_{64}Co_{10}$ thin film.** a) Out-of-plane magnetic hysteresis loops at 175 K (black), at 355 K (green), and at 300 K (red and blue). The red loop corresponds to samples initialized under 355 K and +30 kOe, while the blue loop for 175 K and +30 kOe. b-c) AHE and MR measurements of the 50-μm Hall bar at 355 K (green) and 300 K (red and blue). The red and blue color indicate the same initialization conditions as (a). Arrow pairs are sketched side by side in (b-c) depicting magnetic moment orientations. The inset of (b) shows an example of the magnetic configuration. The left pair indicates the near-compensated Phase I ($M_{Tb}^{I}$ and $M_{FeCo}^{I}$), and the right for the uncompensated Phase II ($M_{Tb}^{II}$ and $M_{FeCo}^{II}$). In each pair the purple arrow represents $M_{FeCo}$ and the orange for $M_{Tb}$. Dash lines are sketched in (b-c) to indicate the major loop enveloping the two biased loops.

To further exploit the EB effect, the magneto-transport behaviors were characterized on the a-TbFeCo Hall bar devices.[23] Theoretically, the anomalous Hall effect (AHE) can be expressed as $R_H \propto R_{Tb}M_{Tb} + R_{FeCo}M_{FeCo}$, where $R_{Tb}$ and $R_{FeCo}$ are AHE coefficients, and $M_{Tb}$ and $M_{FeCo}$ are magnetizations.[37] It is known that in amorphous FM materials, $R_{Tb}$ is positive, while $R_{FeCo}$ is negative.[38] Since $M_{Tb}$ and $M_{FeCo}$ also have opposite signs, the AHE terms of Tb and FeCo actually contribute with the same sign, unlike their compensated contributions to magnetic hysteresis loops. Figure 3(b) shows the AHE loops at 300 and 355 K. The loop at 355 K orients opposite to the magnetic hysteresis loop, because the dominant $M_{FeCo}$ has negative $R_{FeCo}$. Similarly, two exchange biased AHE loops are detected at 300 K, correlated to the two above-mentioned initialization conditions. Moreover, the two biased AHE loops shift away from each other along the $R_H$ axis.

Finally, the transverse magneto-resistance (MR) has been measured using the four-point probe method. The magnetic field was applied perpendicular to the film plane. The current was applied in the film plane and perpendicular to the magnetic field. As shown in Figure 3(c), at 355 K, two sharp antisymmetric peaks were observed in the coercive fields of the corresponding AHE loop. Similar results have been reported in other PMA systems, e.g. Pt/Co multilayers.[39] This type of MR peak is related to the multi-domain configuration during the magnetization reversal process. In addition to the sharp antisymmetric peaks, unusual biased MR loops were revealed at 300 K. The MR difference for the Hall bar of 50-µm width is about 0.1 Ω with a relative change of 0.1 %. These biased MR loops have the same $H_E$ and sample initialization dependence as the AHE and magnetic hysteresis loops. It implies that the biased MR loops are also associated with the EB effect. After all, it should be noted that a larger MR difference (0.2 %) with greater $H_E$ (±6.4 kOe) has been achieved in the amorphous $Tb_{20}Sm_{15}Fe_{55}Co_{10}$ thin films. The details are provided in the supplementary material.[30]

The EB effect can be interpreted by the presence of two nanoscale magnetic phases. Based on the STEM-EDS and APT results, there are two coexisting nanoscale amorphous phases. In the Fe-enriched phase (Phase II) the FeCo moment prevails at room temperature making Phase II behave in a FM manner. Meanwhile, the other phase (Phase I) with higher Tb content provides a near-compensated FiM component. The magnetization and anomalous Hall resistance can be expressed in terms of the contributions from the two phases respectively.

$$M = \phi\left(M_{Tb}^I + M_{FeCo}^I\right) + (1 - \phi)\left(M_{Tb}^{II} + M_{FeCo}^{II}\right), \tag{1}$$

$$R_H \propto C^I\left(R_{Tb}^I M_{Tb}^I + R_{FeCo}^I M_{FeCo}^I\right) + C^{II}\left(R_{Tb}^{II} M_{Tb}^{II} + R_{FeCo}^{II} M_{FeCo}^{II}\right), \tag{2}$$

Where the superscript I and II denote the two nanoscale phases, and $\phi$ is the volume concentration of Phase I. $C^I$ and $C^{II}$ are positive constants related to $\phi$ and conductivity tensor of each phase.[40] Additionally, $M_{Tb}$ is opposite to $M_{FeCo}$ due to the AFM coupling.

The positive and negative $H_E$ of the EB effect depends on the magnetic orientation of Phase I. Specifically, the initialization dependence can be understood by the following discussion. In +30 kOe at 175 K (or 355 K), both $M_{FeCo}^I$ and $M_{FeCo}^{II}$ are aligned to negative (or positive) orientation. As the temperature returns to 300 K, the orientation of Phase I persists and becomes fixed because of its large $H_C$ near compensation. In this way, two distinct orientations of Phase I can be initialized, corresponding to the two opposite biased hysteresis loops. Since all temperatures in this study are essentially lower than the Curie temperature, both nanoscale phases are magnetically ordered, making the initializing process different from the zero-field cooling and field cooling in FM/AFM systems.[41] The cooling or heating process shifts the near-compensated Phase I away from compensation and reduces its coercivity. The critical step here is to align the FiM phase by directly applying a field larger than the reduced coercivity. It has been verified that cooling or heating with zero or non-zero field has an identical effect.

In Figure 3(b-c), the magnetic states of Phase I and II are depicted by colored arrows. The left pair indicates the near-compensated Phase I ($M_{Tb}^I$ and $M_{FeCo}^I$), and the right for the uncompensated Phase II ($M_{Tb}^{II}$ and $M_{FeCo}^{II}$). In each pair the purple arrow represents $M_{FeCo}$ and the orange for $M_{Tb}$. Each biased hysteresis loop contains two magnetic states i.e. parallel and antiparallel states in terms of the relative orientations of $M_{FeCo}^I$ and $M_{FeCo}^{II}$. The antiparallel state is metastable due to its higher exchange energy than the parallel state. Thus, it requires a larger field to switch from the parallel state to the antiparallel state and a smaller field the other way around, making the hysteresis loop exchange biased. Based on Equation (1) and (2), the biased loops also shift along the $M$ and $R_H$ axes because of the fixed contribution from Phase I. Since Phase I is near-compensated, $M_{Tb}^I + M_{FeCo}^I$ is close to zero. So the $M$-shift of the magnetic hysteresis loop is very small along $M_{FeCo}^I$ above $T_{comp}$. On the other hand, as discussed above, $C^I(R^I_{Tb}M^I_{Tb} + R^I_{FeCo}M^I_{FeCo})$ always provides a finite contribution, resulting in a finite $R_H$-shift opposite to $M^I_{FeCo}$. It is noteworthy that positive EB together with upward $M$-shift has been reported in the Fe/FeF$_2$ system with an AFM interfacial coupling.[42] However, in our study the positive EB corresponds to a tiny downward $M$-shift as well as an upward $R_H$-shift. This fact implies that the overall interfacial coupling at 300 K is FM, consistent with the larger FeCo contribution in both phases.

In addition, the shape of the biased loops is asymmetric, especially of the biased AHE loops. The metastable antiparallel state gradually evolves (or "rotates") to the stable parallel state on the other loop. This transition indicates a big major loop with greater coercive field enveloping both biased loops and connecting the two isolated branches. It should be noted that the observed exchange bias is a minor loop effect. The major loop would be symmetric if sufficiently large fields were applied to switch Phase I. Moreover, the temperature has a significant effect on the EB as well. At high temperature, the magnetic ordering mainly comes from the FeCo spins making both of the phases FM dominated. On the other hand, at low temperature, the two phases magnetically merge into a rigid FiM phase because of the large Tb atomic moment and single-ion anisotropy. Thus, the EB vanishes at both low and high temperature. Finally, further magneto-transport modeling of the two nanoscale phases is necessary to lead to a deeper understanding of the underlying physics.

The model of two nanoscale phases implies the existence of the bistable MR states. The magnetic states are depicted along with the biased MR loops in Figure 3(c). Obviously, the MR value mainly depends on the relative orientation of the two nanoscale phases. The parallel states have lower MR than the antiparallel states similar to the tunnel MR in magnetic tunnel junctions. In this way we obtain a pair of MR states on each exchange biased MR loop. Moreover, the MR value can be switched between these states by magnetic field impulses, and the states have been proved to be stable at room temperature. The results of switching and stability study are provided in the supplementary material.[30]

In summary, this work provides evidence of the EB and bistable MR states in the a-TbFeCo thin films. The EB is closely related to the two growth-induced nanoscale phases distributed throughout the film, which were observed using STEM-EDS and APT. This biased thin film has many appealing properties such as the large PMA and room-temperature capability. Moreover, the amorphous thin film requires no epitaxial growth and specific substrates. The bistable MR states associated with the EB are proved to be stable at room temperature and switchable by sweeping the magnetic field. More tunability of the EB and bistable MR states can be achieved with a wide range of alloy compositions and controllable growth of nanoscale phases. Further efforts are under progress to vary the sputter deposition parameters and study the tuning of magnetic properties as a function of deposition conditions. Recently, all-optical switching (AOS) using ultrafast laser has been discovered in a-TbFeCo thin films.[5, 43] Our findings also imply that the biased a-TbFeCo thin film may be potential for the new AOS devices.

*Acknowledgements*

The work at University of Virginia was partially supported by the Defense Threat Reduction Agency grant (Award No. HDTRA 1-11-1-0024). Atom probe tomography was performed using Environmental Molecular Sciences Laboratory (EMSL), a national scientific user facility sponsored by the Department of Energy's Office of Biological and Environmental Research. EMSL is located at PNNL, a multi-program national laboratory operated by Battelle Memorial Institute under Contract No. DE-AC05-76RL01830 for the U.S. Department of Energy. A.D. would like to acknowledge the funding from the Material Synthesis and Simulations Across Scales (MS$^3$) Initiative conducted under the Laboratory Directed Research and Development Program at Pacific Northwest National Lab (PNNL). R.B.C would like to acknowledge support from the Linus Pauling Distinguished Postdoctoral Fellowship through the Laboratory Directed Research and Development Program at PNNL.